\documentclass[pra, twocolumn,showpacs]{revtex4}

\usepackage[dvips]{graphicx}

\begin{document}

\title{Alignment dynamics of slow light diffusion in ultracold atomic $^{85}$Rb}

\author{S. Balik, R.G. Olave, C.I. Sukenik, and M.D. Havey}
\affiliation{Department of Physics, Old Dominion University,
Norfolk, VA 23529}

\author{V.M. Datsyuk, I.M. Sokolov, and D.V. Kupriyanov}
\affiliation{Department of Theoretical Physics, State Polytechnic
University, 195251, St.-Petersburg, Russia}

\date{\today}

\begin{abstract}
A combined experimental and theoretical investigation of time- and
alignment-dependent propagation of light in an ultracold atomic
gas of atomic $^{85}$Rb is reported.  Coherences among the
scattering amplitudes for light scattering off excited hyperfine
levels produce strong variations of the light polarization in the
vicinity of atomic resonance. Measurements are in excellent
agreement with Monte-Carlo simulations of the multiple scattering
process.
\end{abstract}

\pacs{32.80.-t, 32.80.Pj, 34.80.Qb, 42.50.-p, 42.50.Gy, 42.50.Nn}

\maketitle Disordered systems have been considered too complex for
research into fundamental properties of physical systems. However,
technical advances in creation and manipulation of coherence in
mesoscopic samples, such as quantum degenerate gases, have made
the influence of disorder in atomic and in condensed phases of
considerable interest.  The essential role disorder can play in
phase transitions was pointed out by Anderson \cite{Anderson} in
consideration of localization of electrons by disorder. More
recent research has focused on general localization phenomena,
including localization of matter waves by random or quasi random
optical lattices \cite{Damski,Krug}.  In this case, the resulting
Anderson and Bose glass phases represent a phase transition in the
transport properties of the matter waves by optical disorder.

Another research focus area has been localization of
electromagnetic waves in strongly scattering and disordered
condensed or atomic media \cite{Sheng,LagTig,Wiersma1,Chabanov1}.
Previous studies have considered only massive particles and, in
comparison, light localization seems a fundamentally different
phenomenon. Nevertheless, there have been two intriguing reports
of light localization in condensed samples consisting of classical
scatterers \cite{Wiersma1,Chabanov1}. With characteristic strong
and narrow scattering resonances and well-known interactions with
light and external static fields, ultracold atomic gases are being
considered as possible systems in which to study light
localization.  To attain localization, it is generally believed
that the Ioffe-Regel condition, $\emph{kl}\leq1$, must be
satisfied \cite{Sheng}. Here $k$ is the local wave vector of the
light, and $\emph{l}$ is the scattering mean free path. For near
resonance scattering in an atomic gas, this implies a
$\lambda^{-3}$ density scaling, giving a required density larger
than $10^{13}$ atoms/cm$^{3}$. Physically this means that the
light scattering is in the near-field regime. Among other
techniques, recent developments in all-optical approaches for
forming and manipulating ultracold gases have achieved densities
in this range \cite{Barrett}. Light localization studies typically
distinguish two limiting cases, one being the strong localization
regime where the Ioffe-Regel condition is satisfied. In the weak
localization, lower density limit, $\emph{kl} \gg 1$. Then, for
non quantum degenerate atomic gases, scattering may be thought of
as a sequence of scattering and propagation events. However, for
ultracold gases in the weak localization regime, quantum
interference plays an important role in light transport. First
experiments on interference effects in multiple light scattering
in ultracold atomic gases were the measurements, in ultracold
atomic $^{85}$Rb, of coherent backscattering by Labeyrie, \emph{et
al.} \cite{Labeyrie1}. In coherent backscattering an
interferometric enhancement of the intensity of scattered light is
measured in a narrow cone in the nearly backwards direction. The
enhancement comes about because reciprocal scattering paths within
the medium have phase relations that survive configuration
averaging. The experiment was important because it demonstrated
breakdown of classical description of light transport in an atomic
vapor. It also stimulated studies demonstrating novel interference
phenomena associated with magnetic \cite{Magnetic}, nonlinear
optical processes \cite{Nonlinear} and hyperfine interferences
\cite{Antilocalization}.

Although an important goal is to achieve strong light localization
in an ultracold atomic gas, all multiple scattering experiments to
date have been done in the weak localization limit. The
experiments have clarified a number of unique features of multiple
light scattering in ultracold atomic gases. For example, recent
experiments \cite{LabeyrieTime,Fioretti} reported on the
time-evolution of light scattered from optically thick samples of
ultracold alkali metal atoms.  In the experiments of Ref.
\cite{LabeyrieTime} in $^{85}$Rb, attention was focused on the
delay time associated with the process, which consists of a
transport and a dwell time \cite{LagTig,Muller}.  The combination
was shown, in a range on the order of the natural width of the
transition, to be independent of detuning. This important result
demonstrated the essential roles that scattering and transport
processes play in the time scale for light transport. The
experiment reported a small diffusive energy velocity $\sim$
$10^{-5}c$, where $c$ is the vacuum speed of light. Note that this
is not the more familiar slow-light behavior observable in the
coherent beam due to electromagnetically induced transparency
\cite{Slowlight}. The time scale here describes incoherent flow of
energy through the medium. The experiments of Ref.
\cite{LabeyrieTime} however, did not discuss the contribution, to
the multiple scattering dynamics, of the atomic alignment produced
in atomic excitation with polarized light. In the studies reported
here, we have determined the dynamics of the atomic alignment
produced in an ultracold gas of $^{85}$Rb under conditions similar
to those reported in \cite{LabeyrieTime}. Measurements include
observation of spectral variations of the alignment in a range of
several natural widths ($\gamma$) around the atomic resonance, and
of the time evolution and polarization of the multiply scattered
light. The data are compared with Monte-Carlo simulations of the
processes, and found to be in good quantitative agreement. The
most important result is that interferences among hyperfine
scattering amplitudes strongly influence light propagation in the
atomic gas, and may significantly impact efforts to obtain strong
localization in an optically dense atomic sample.

\begin{figure}[tp]
\includegraphics{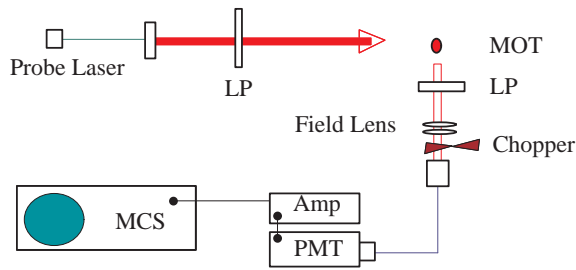}
\caption{A schematic diagram of the experimental arrangement.
Shown is a magneto optical trap (MOT), linear polarizers (LP), and
a photomultiplier tube (PMT). Data is time-binned and stored in
the multichannel scalar (MCS).}
\label{Fig.1}%
\end{figure}%

As in Fig. 1, the experiment is performed on an ultracold gas of
atomic $^{85}$Rb prepared in a magneto optical trap operating on
the $F = 3 \rightarrow F' = 4$ hyperfine transition. The trap,
which has been described elsewhere \cite{HaveyCBS}, produces a
nearly Gaussian cloud of $\sim$ $10^{8}$ ultracold rubidium atoms
at a temperature $\sim$ 100 $\mu$K. The peak density is $\sim$ $3$
x $10^{10}$ $cm^{-3}$. The Gaussian radius of the sample is
$r_{0}$ $\sim$ $1 mm$, determined by fluorescence imaging.
Measurement of the spectral variation of the transmitted light
gives a peak optical depth of $b_{0}$ = 8(1). For a Gaussian atom
distribution in the trap, the maximum weak-field optical depth is
given by $b_{0}$ = $\sqrt{2\pi}$$n_0$$\sigma_{0}$$r_{0}$. Here
$n_0$ is the peak trap density and $\sigma_{0}$ is the
on-resonance cross-section. The isolated-resonance scattering
cross section $\sigma$ varies with probe frequency, $b = b_{0}[1 +
(2\Delta/\gamma)^{2}]^{-1}$, where $\Delta = \omega_{L} -
\omega_{0}$, and $\omega_{L}$ is the probe frequency, $\omega_{0}$
is the $F = 3 \rightarrow F' = 4$ hyperfine transition frequency.
A weak probe laser is tuned in a range of several $\gamma$ around
this transition. The laser is a continuous wave diode laser having
a bandwidth $\sim$ 1 MHz, and an average light intensity of 1 $\mu
W/cm^{2}$. To produce a nearly Gaussian beam profile, the laser
output is passed through a single-mode optical fiber.  The beam is
then expanded and collimated to a $1/e^{2}$ width $\sim$ 8 mm. The
probe laser intensity is modulated with an acousto optic modulator
(AOM), which generates nearly rectangular pulses having an
\emph{on} time of 2 $\mu s$ and an \emph{off} time of 2 $ms$. The
2 $\mu s$ excitation pulse is centered in a 90 $\mu s$ window
during which fluorescence signals are recorded. The MOT lasers are
off during this period. For the remaining nearly 2 $ms$, the MOT
lasers are turned back on to reconstitute the atomic sample.
Fluorescence from the MOT region present during this period is
prevented from reaching the PMT by a synchronized mechanical
chopper. The AOM-limited 20 dB response is $\sim$ 60 ns. The probe
laser is vertically polarized.

Scattered light signals are detected in a direction orthogonal to
the probe laser propagation and polarization directions.  The
light is collected in a solid angle of about 0.35 mrad, and
refocussed to match the numerical aperture of a 400 $\mu$m
multimode fiber. A linear polarization analyzer is placed between
the MOT and the field lens to collect signals in orthogonal linear
polarization channels, labelled as parallel ($\|$) and
perpendicular ($\bot$). The polarization response is calibrated
against the known polarization direction of the probe laser; the
measured 20 percent difference in sensitivity is used to correct
the signals taken in the two channels. The fiber output is coupled
through a 780 nm (5 nm spectral width) interference filter to a
GaAs-cathode photomultiplier tube (PMT). The PMT output is
amplified and directed to a discriminator and multichannel scalar,
which serves to sort and accumulate data into 5 ns bins. A
precision pulse generator is used to control timing of the MOT and
probe lasers and for multichannel scalar triggering.

\begin{figure}[tp]
\includegraphics{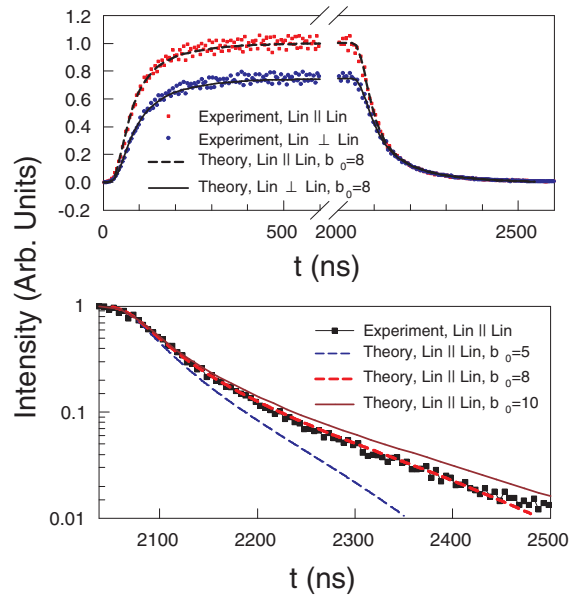}
\caption{Measured time-dependent scattered light signal in
orthogonal $P_{L}$ channels. Theoretical results are indicated by
the curves, as labelled in the figure legend. The peak intensity
corresponds to about 10$^{4}$ counts.}
\label{Fig.2}%
\end{figure}

The measured intensities in two orthogonal polarization channels
for resonance excitation of the $F = 3 \rightarrow F' = 4$
hyperfine transition are shown in Fig. 2.  Note that the peak
intensity in the lin $\|$ lin channel is very nearly 10${^4}$
counts, which corresponds to 10 experimental runs, each with a 120
s data accumulation period. The time response of the data
acquisition system, including the AOM switching, is fast in
comparison with the time evolution of the fluorescence signals. We
point out that the transient build up of several hundred ns is due
to multiple scattering of probe radiation after it is switched on.
The time scale for the process can be seen more clearly in the
lower panel of Fig. 2. The first 50 ns of this curve is distorted
by the electronic shutoff of the probe pulse.  Beyond that, the
decay curve is multiexponential, and varies from the natural
single atom fluorescence from atoms located near the surface of
the sample to longer-time-scale decay arising from atoms deeper
within the sample. The solid curves in Fig. 2 represent
Monte-Carlo simulations of the scattering process. Other than the
overall intensity scale, there are no adjustable parameters in the
comparison, with the simulation input data consisting of the
measured trap density profile and the AOM response. The agreement
is excellent, showing that the physics of the process is well
modelled.

The fluorescence time behavior given in Fig. 2 suggests for longer
times an approximately exponential decay, with estimated time
constant of 170(20) ns. It is generally expected that the longest
time scale reflects the sample geometry, and is given by a single
exponential, often termed the lowest-order Holstein mode
\cite{Holstein}. In this regard, our results are in qualitatively
good agreement with the those of \cite{LabeyrieTime} for our
optical depth b $\sim$ 8(1). In \cite{LabeyrieTime}, it is also
shown that the longest decay time, in an elastic diffusion theory
and for large optical depth, scales for a Gaussian atom
distribution, as $\tau_{0} = 0.057\tau_{nat}b^{2}$, where
$\tau_{nat}$ = 27 ns is the natural decay time of the excited
level. Although this result gives qualitatively good agreement
with experiment, it seems to underestimate, at lower optical
depths, the measured decay time, both in our results and in those
of \cite{LabeyrieTime}. This difference may be due to departure of
our atomic sample from an ideal Gaussian atom distribution, or to
approximations made in the boundary conditions of the diffusion
model \cite{Sheng}.

\begin{figure}[tp]
\includegraphics{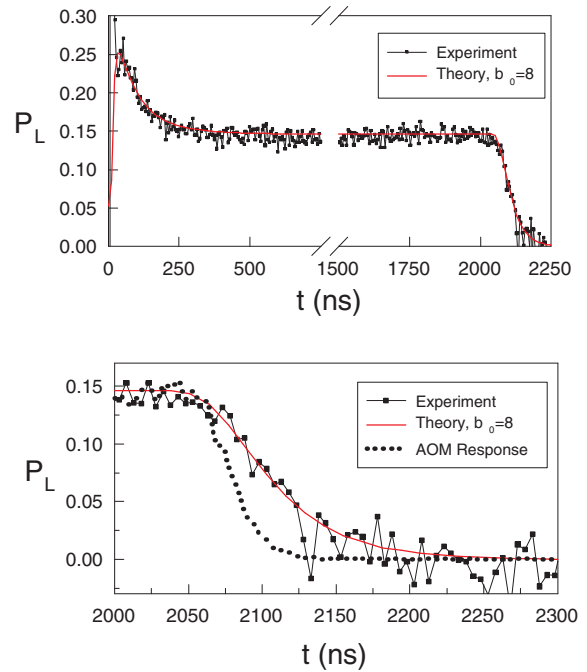}
\caption{Measured time-dependent $P_{L}$, shown as solid points.
Monte Carlo simulation results are shown as solid lines, while the
limiting AOM response is shown as a dotted line.}
\label{Fig.3}%
\end{figure}

From Fig. 2 it is also clear that the fluorescence signals are
different for the lin $\|$ lin and lin $\bot$ lin polarization
channels. This effect is quantified by defining a linear
polarization degree as
\begin{equation}
P_{L} = \frac{I_{\parallel} - I_{\perp}}{I_{\parallel} +
I_{\perp}} = \frac{-15\langle A_{0}\rangle}%
{28-5\langle A_{0}\rangle},%
\label{PL}%
\end{equation}
In the formula, $I_{\parallel}$ and $I_{\perp}$ represent the
measured intensities in the lin $\parallel$ lin and lin $\perp$
lin channels. We emphasize that $P_{L}$ is related to the
electronic alignment generated by excitation of an initially
unpolarized atomic gas of ground state atoms with linearly
polarized light. Then only the average axially symmetric alignment
component $\langle A_{0}\rangle$ is nonzero. The alignment is
defined in terms of the upper state hyperfine angular momentum
operators as the ensemble average $\langle A_{0}\rangle = \langle
3\hat{F'}_{z}^{2} - \hat{\mathbf{F'}}^{2}\rangle/F'(F'+1)$. In
$P_{L}$ above, the expression in terms of $\langle A_{0}\rangle$
is correct for small detunings from the resonance line, where
contributions from the $F = 3 \rightarrow F' = 2, 3 \rightarrow F
= 3$ transitions may be ignored.  Finally, we point out that the
above discussion ignores inelastic Raman transitions to the lower
$F = 2$ hyperfine level, which have a negligible effect on the
reported data. The data in Fig. 2 give the time-dependence of
$P_{L}$ shown in Fig. 3. There we see that $P_{L}$ enhances the
differences in the two channels, showing the time-dependent
maximum in $P_{L}$ soon after the exciting pulse is turned on.
This is followed by approach to a steady state $P_{L}$, which
decays rapidly upon switching off the exciting laser pulse. The
varied behavior can be understood by considering that when the
exciting laser is first turned on, the prompt signal comes mainly
from single scattering events. Then the peak value of $P_{L}$
should be close to the single scattering value of $P_{L}$ = 0.268,
as is seen in Fig. 3. Second, even though the light scattering is
nearly elastic, the polarization state of the scattered light will
be randomized in multiple scattering by the presence of the
multiplicity of available elastic Raman and Rayleigh radiative
channels. Note that inelastic Raman transitions to the lower $F =
2$ level are negligible in the spectral range of data reported
here. Then we expect (and observe) that the steady state
polarization, which has contributions from multiple order
scattering events, is lower than the single scattering value, but
is still nonzero. However, as seen in the lower panel of Fig. 3,
once the exciting laser is turned off, $P_{L}$ rapidly decays to a
very small value. Note that the AOM-determined shut off time for
the exciting light is not negligible on the scale of the $P_{L}$
decay. However, it is clear that, after a few atomic radiative
lifetimes, $P_{L}$ has decayed to a small value. The decay is
monotonic, and roughly exponential, with a decay constant on the
order of two natural lifetimes. This is to be contrasted with the
decay of the total excitation (see Fig. 2), where population
survives for much longer time scales. The alignment generated by
optical excitation is then fragile in comparison with the
population.  Finally, theoretical results
\cite{LaserPhysicsReview} for the decay rate of $P_{L}$ as a
function of $b_0$ show that $P_{L}$ decays more slowly as the
optical depth is reduced. This is physically plausible as, in a
single scattering limit, $P_{L}$ would remain constant as the
atomic population decayed.

\begin{figure}[tp]
\includegraphics{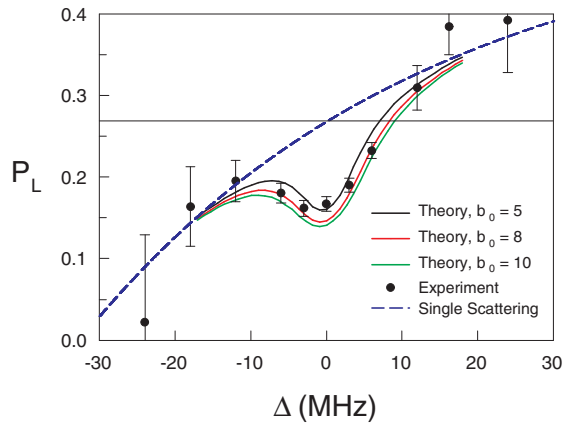}
\caption{Detuning dependence of $P_{L}$. The dots represent
experimental data points, while the blue chained curve is the
expected variation for single atom scattering. The red and blue
curves indicate variation for optical depths of $b_0 = 5$ (black),
$b_0 = 8$ (red), and $b_0 = 10$ (green).}
\label{Fig.4}%
\end{figure}%

As shown in Fig. 4, we have recorded data for a detuning range of
$\pm$ 24 MHz around the $F = 3 \rightarrow F' = 4$ hyperfine
transition. We see that, as the magnitude of the detuning is made
larger, the measured steady-state values for $P_{L}$ approach the
calculated frequency-dependent single scattering limit. This limit
varies with detuning because of interference of the scattering
amplitudes among the $F = 3 \rightarrow F' = 2,3,4$ hyperfine
transitions, and does not depend on the existence of multiple
hyperfine levels in the 5s $^{2}S_{1/2}$ lower energy level. In
the absence of interference, $P_{L}$ is nearly constant, as
indicated by the horizontal line in Fig. 4. The single scattering
curve is expected, for as the laser is tuned away from resonance,
the optical depth decreases, and the number of contributing
scattering orders decreases. The solid curves in Fig. 4 represent
theoretical results for different optical depths $b_0$. We see
that the experimental data (with one sigma error bars) is
bracketed by the $b_0$ = 5 and $b_0$ = 8 calculations, consistent
with the experimental on-resonance optical depth of $b_0 = 8 (1)$.

In conclusion, we have reported frequency, time and
polarization-dependent measurements of near-resonance fluorescence
emitted, in a multiple scattering regime, from ultracold atomic
$^{85}$Rb atoms. The measurements are in excellent agreement with
Monte-Carlo simulations of the process.  The results show that
light scattered from the atomic ensemble maintains, for a time
scale of several atomic lifetimes, a residual of the initial
electronic alignment created by excitation with linearly polarized
light. The steady state polarization varies strongly in the
vicinity of atomic resonance, demonstrating that hyperfine
interferences in the scattering amplitudes play a critical role in
light transport in a dense atomic gas.

Supported by the National Science Foundation (NSF-PHY-0355024),
the Russian Foundation for Basic Research (RFBR-05-02-16172-a),
and the North Atlantic Treaty Organization (PST-CLG-978468).

\end{document}